\documentclass[10pt,twocolumn,a4paper]{article}

\usepackage[left=20mm, right=15mm, top=15mm, bottom=15mm]{geometry}

\usepackage{subfig}
\usepackage{flushend}
\usepackage{indentfirst}
\usepackage{graphics}
\usepackage{amsmath}
\usepackage{graphicx}
\usepackage{epstopdf}
\usepackage{float}

\graphicspath{{./figure/}}

\begin{document}

\title{\huge \textbf{A Toy Model of Complete Cosmic History}}

\date{}

\twocolumn[
\begin{@twocolumnfalse}
\maketitle

\author{\textbf{V. K. Shchigolev}$^{1,*}$\\\\
\footnotesize $^{1}${Department of Theoretical Physics, Ulyanovsk State University, 42 L. Tolstoy Str., Ulyanovsk 432000, Russia}\\

\footnotesize $^{*}$Corresponding Author: vkshch@yahoo.com}\\\\\\

\end{@twocolumnfalse}
]

\noindent \textbf{\large{Abstract}} \hspace{2pt} In the present paper, we study a  toy cosmological model derived from the specific behavior of the Hubble parameter and the scale factor in a spatially-flat Friedmann-Robertson-Walker (FRW) space-time. We demonstrate that our model could match in some approximation the complete history of cosmic expansion. To establish the appropriate  values of the the model parameters, that is to fit the real universe, we apply some theoretical and observational tests. \\

\noindent \textbf{\large{Keywords}} \hspace{2pt} Cosmology, Toy Model, Cosmic History, Accelerated Expansion, Theoretical and Observational Tests, Scalar Field\\

\noindent\hrulefill

\section{\Large{Introduction}}

Present accelerated expansion of the universe is well
proved in many papers [1--9]. In order to explain so
unexpected behavior of our universe, one can modify the
gravitational theory, or construct various
field models of the so-called dark energy (DE) which equation of state
satisfies $w= p/\rho< -1/3$.
The simplest candidate of DE is the cosmological constant
with $ w = -1$.  If it is quintessence then $-1 < w <- 1/3$ and if it is phantom
then $ w <-1$. The constant  equation of state  $w = -1$  is called phantom divide.

So far, a large class of scalar-field dark energy models have
been studied, including tachyon, ghost condensate
 and quintom, and so forth. In addition, other
proposals on DE include interacting DE models,
braneworld models, and holographic DE models, etc. There are some dark energies which can
cross the phantom divide from both sides. Nevertheless, the puzzle of  dark energy is still an unsolved problem in modern cosmology (see, e.g. \cite{Miao} and references therein).

At the same time, we firmly believe in the paradigm of cosmological inflation regarding the very early universe. If this is so, any cosmological models should have at least two periods of accelerated expansion: early and late. It is hard to imagine a model that would be able to describe the extremely difficult scenario for the evolution of the universe from the origin to the present day, with all its processes and content. A variety of models of the processes in the early and late universe calls into question even perspective obtaining of such models in the near future.
In this respect, it is interesting to consider any analytic model, even a toy model, capable of, even superficially, but realistically describe the whole evolution from (almost) the beginning to the present day (see, e.g. [11--16]).

We briefly study a model in which the Hubble parameter $H$ takes a specific form, providing two accelerated epoch in the evolution of the universe. In this regard, we
consider a spatially flat Friedmann-Robertson-Walker background equipped with such a time-varying $H(t)$.  To testify this model, we consider some theoretical and observational tests. In order to simplify our study, we consider only a limited number of those tests. It should be noted all that tests could essentially improve any models (see, for example,  \cite{Darabi}) and bring them closer to the real scenario of cosmic evolution.

\section{\Large{The Model}}

The Einstein's field equations without a cosmological $\Lambda$ - term can be written  as
\begin{equation}\label{1}
R_{ik}- \frac{1}{2} g_{ik} R = T_{ik},
\end{equation}
where  we assume  that the gravitational constant $8\pi G=1$, and all symbols have their usual meanings in the Riemannian geometry.
The energy-momentum tensor of matter $T_{ik}$ can be derived in a usual manner from the Lagrangian of matter. Considering the matter to be a perfect fluid with the energy density $\rho$ and pressure $p$ , we have
\begin{equation}\label{2}
T_{ik}= (\rho +p)u_i u_k -p\, g_{ik},
\end{equation}
where  $u_i = (1,0,0,0)$ is  4-velocity of the co-moving observer, satisfying $u_i u^i = 1$.

The line element of a Friedmann-Robertson-Walker (FRW) is represented by
$$ds^2 = d t^2- a^2 (t)\Big[d r^2+\xi^2 (r)d \Omega^2\Big],$$
where $a(t)$ is a scale factor of the Universe, $\xi(r)=\sin r,r,\sinh r$ in accordance with the  sign of spatial  curvature $k=+1,0,-1$.
Given this metric and (\ref{2}), we can reduce the field equation (\ref{1}) to the following set of equations:
\begin{eqnarray}
3H^2 + \frac{3 k}{a^2} = \rho,\label{3}\\
2 \dot H + 3H^2 + \frac{k}{a^2}  = -  p,\label{4}
\end{eqnarray}
where $H = \dot a/a $ is the Hubble parameter, and an overdot stands for  differentiation with respect to cosmic time $t$.

The continuity equation follows (\ref{3}) and (\ref{4}) as:
\begin{equation}\label{5}
\dot \rho + 3 H (\rho + p)=0.
\end{equation}

\subsection{\normalsize{A spatially flat FRW cosmology}}

From now on, we consider a spatially flat FRW cosmology with $k=0$.
In this case, one can rearrange the basic equations of the model, (\ref{3}) and (\ref{4}), as follows:
\begin{equation}\label{6}
3H^2  = \rho,
\end{equation}
\begin{equation}\label{7}
2 \dot H  = - \rho - p.
\end{equation}
As can be easily verified, the continuity equation (\ref{5}) follows from the set of equations (\ref{6}), (\ref{7}). So, we have only two independent equations in the set (\ref{5})-(\ref{7}).
The generic procedure for solving this system lies in  specification of the material or sources of gravitational field, possessing the energy density $\rho$ and pressure $p=w\rho$. Here, the equation of state parameter $w$ (the barotropic index) can be a constant or a function of time.

On the other hand, two equations, for example (\ref{6}) and (\ref{7}), are quite enough to find two unknown parameter such as $\rho$ and $p$, if we specify the Hubble parameter $H$ as a function of time. Let us consider one such a law for $H(t)$, which is appeared in other context in  \cite{1Shchigolev,2Shchigolev}. This law is similar, but not coincide with  the results suggested in [20--22], and can be written as follows
\begin{equation}\label{8}
H(t)=H_0\left(1+\frac{m n}{(H_0 t)^{\displaystyle n+1}}\right).
\end{equation}
where $m,\,n$ are some positive dimensionless constants, and $H_0$ is the asymptotic (at large cosmic time) Hubble constant.  Making use of this equation, we can obtain the time evolution of the scale factor represented by
\begin{equation}\label{9}
a(t)=a_0 \exp\Big[H_0 t -\frac{m}{(H_0 t)^{\displaystyle n}}\Big],
\end{equation}
where $a_0$ is the scale factor at instant $t_0=m^{\displaystyle 1/(n+1)}/H_0$. In order of magnitude, it coincides with the age of the universe $t_0 \approx 13.7\times10^9 \,yrs=4.32\times 10^{17}\, s$. Therefore, the constant $a_0$ is approximately the present value of scale factor.

In Fig. 1, we show the time evolution of the scale factor and the Hubble parameter, for three deferent values of $n$. Corresponding values of $m$ are chosen from the condition $w_{max}=1/3$ as shown below. It is interesting to note that the so-called hybrid expansion law is a kind of the limiting case in (\ref{8}): $n \to 0$ simultaneously with  $mn \to M =constant \neq 0$ (see, e.g. \cite{2Shchigolev,Akarsu}).

Substituting (\ref{8}) into (\ref{6}) and (\ref{7}), we can find both $\rho$ and $p$ as the functions of time:
\begin{equation}\label{10}
\rho = 3H_0^2\left(1+\frac{m n}{\tau^{\displaystyle n+1}}\right)^2,
\end{equation}
\begin{equation}\label{11}
p = \frac{H_0^2}{\tau^{\displaystyle n+2}}\left[2mn(n+1)-\frac{3}{\tau^{\displaystyle n}}\left(m n +\tau^{\displaystyle n+1}\right)^2\right],
\end{equation}
where $\tau = H_0 t$ is the dimensionless  cosmic time.
\begin{figure}[thbp]
\centering
\includegraphics[width=0.45\textwidth,height=0.4\textwidth]{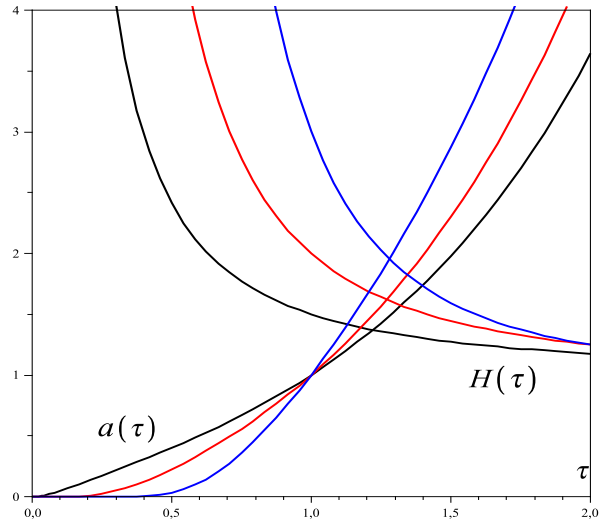}
\caption{Graphs of the scale factor $a(\tau)$ and the Hubble parameter $H(\tau)/H_0$ versus time $\tau$. The black lines are plotted for $n=1/2,\,m=(25/96)\sqrt{5/3}$, the red lines correspond to $n=1,\,m=27/256$, and the blue lines are of the case $n=2,\,m=1/27$.}
\label{Figure_1}
\end{figure}
Thus, the barotropic index $w=p/\rho$ is given by
\begin{equation}\label{12}
w(\tau)=-1 + \frac{2}{3}\,\frac{m n (n+1) \tau ^{\displaystyle  n}}{\Big[m n + \tau ^{\displaystyle n+1}\Big]^2}.
\end{equation}
Evolution of the equation of state in our model according to the last equation for the certain values of parameters is shown in Fig. 2. In order to construct a simple (and yet not completely trivial) model for the complete cosmic history, let us outline the generic features we want to reproduce by means of our model.

It is easy to find that $w$ starts at  $w(\tau=0)=-1$ at the initial time
and asymptotically tend to the same value:  $w(\tau \to \infty)=-1$.
However, during a certain time interval, the equation of state may become positive: $0<w<1$. Furthermore,   the expansion slows down even for a longer period. After that, the expansion again accelerates. It is interesting that this equation of state gains $w=w_{max}>0$ in its maximum. One can find out from (\ref{12}) that
\begin{equation}\label{13}
w_{max} = -1 +\frac{n(n+2)}{6(n+1)}\, \frac{1}{\tau_m}.
\end{equation}
The corresponding instant of reaching this maximum is given by
\begin{equation}\label{14}
\tau_m =\left(\frac{m n^2}{n+2}\right)^{\displaystyle \frac{1}{n+1}}.
\end{equation}
From (\ref{13}) and (\ref{14}), we can derive the magnitude of parameter $m$ which provides this maximum with the given value of $w_{max}$ in it:
\begin{equation}\label{15}
m=\frac{n^{n-1}(n+2)^{n+2}}{\Big[6(n+1)(w_{max}+1)\Big]^{n+1}}.
\end{equation}
\begin{figure}[thbp]
\centering
\includegraphics[width=0.4\textwidth,height=0.35\textwidth]{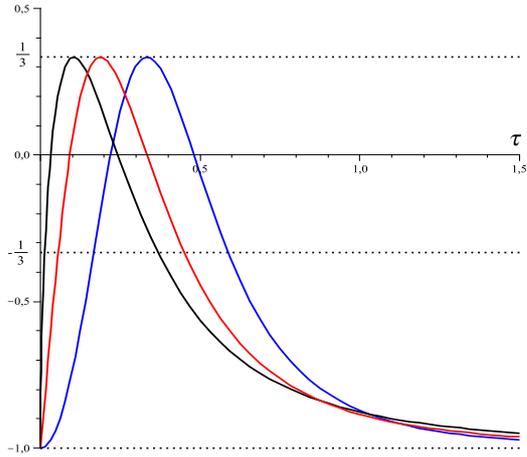}
\caption{Graphs of the equation of state versus time. Here, we use the same convention on values of $n$ and $m$ as in Fig. 1.}
\label{Figure_2}
\end{figure}
In our view, all these features of the model bring it closer to the realistic scenarios, widely discussed at present. As one can see, the tuning of this model is possible by means of several parameters, such as $n,\,m,\, H_0,\,a_0$ and $w_{max}$, and  retains the basic features of this model that makes it so attractive.

To investigate the possibility of accelerated expansion for the universe in the framework of our model, we take into account the so-called "deceleration parameter" $q=-a \ddot a/\dot a^2=-1-\dot H/H^2$. From (\ref{9}), we can obtain
\begin{equation}\label{16}
q(\tau)=-1 + \frac{m n (n+1) \tau ^{\displaystyle  n}}{\Big[m n + \tau ^{\displaystyle n+1}\Big]^2}.
\end{equation}
It should be noted that the time evolution of this deceleration parameter is also clear from Fig. 2 since its graph crosses the line $q=0$ at the same time as $w(\tau)$ crosses the line $w=-1/3$.
Such a behavior of the deceleration parameter indicates the existence of two periods of accelerated expansion in our model, the initial and the final, separated by the period of decelerated expansion. Namely this feature of the model makes it so attractive.

\section{\Large{Testing the Model}}

Here, we want to give several some diagnostics of our model based on theoretical background and on the observational data.

\subsection{\normalsize{Square adiabatic sound speed}}

As well known \cite{Miao}, there are several approaches to investigate stability of a model. For instance, one can use the fact that the sound-speed squared should be non-negative. This parameter is useful to investigate the classical
stability of the models, therefore if sound speed be
equal to a positive quantity the proposal can be considered
as a viable model. The squared adiabatic sound
speed is defined as
$$
c^2_s=\frac{d p}{d \rho}=\frac{\dot p}{\dot \rho}.
$$
Due to (\ref{6}) and (\ref{7}), the stability condition is given by
\begin{equation}\label{17}
c^2_s = -1 - \frac{\ddot H}{3 H \dot H} \ge 0.
\end{equation}

Let us apply this relation to our model. Substituting (\ref{8}) into (\ref{14}), we conclude that this stability condition can be expressed as
\begin{equation}\label{18}
c^2_s=-1+\frac{(n+2)\,\tau^{\displaystyle n}}{3\Big(mn+\tau^{\displaystyle n+1}\Big)} \ge 0.
\end{equation}
\begin{figure}[thbp]
\centering
\includegraphics[width=0.4\textwidth,height=0.35\textwidth]{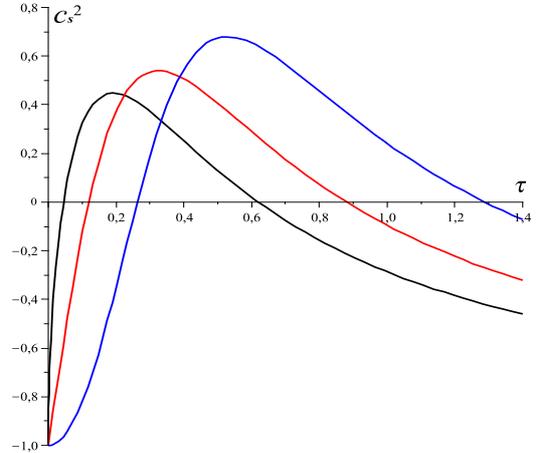}
\caption{Square adiabatic sound speed versus time. We also use the same convention on values of $n$ and $m$ as in Fig. 1.}
\label{Figure_3}
\end{figure}
From Fig. 3 it is clear that there exists a span during which the model is stable in respect to the sound-speed criteria for any values of the model parameters. It is noteworthy that increasing value of $n$ allows the sound-speed condition of stability to be  satisfied during the whole late acceleration period up to the present.

\subsection{\normalsize{Statefinder diagnostic}}

In order to distinguish between  various DE models, V. Sahni et al. \cite{Starobinsky,Alam} proposed a cosmological diagnostic pair $\{r, s\}$ called statefinder.
The statefinder test is a geometrical one based on the expansion of the scale factor a(t) near the present time $t_0$:
$$
a(t) = 1 + H_0 (t - t_0) -\frac{1}{2} q_0 H^2_0 (t - t_0)^2 +\frac{1}{6}r_0 H^3_0 (t - t_0)^3 + ...,
$$
where $a(t_0) = 1$ and $H_0, q_0, r_0$ are the present values of the Hubble parameters, deceleration parameter and the statefinder index $ r = \dddot a /a H^3$ respectively. The statefinder parameter $s$ is the combination of $r$ and $q$: $s = (r - 1)/3(q - 1/2)$.
The important feature of statefinder is that the spatially flat $\Lambda$CDM has a fixed point
$\{r, s\} = \{1, 0\}$. Departure of a DE model from this fixed point is a good way of establishing the 'distance' of this model from flat $\Lambda$CDM. In terms of the Hubble parameter and its derivatives with respect to cosmic time, the statefinder parameters of a flat FRW model are given by
$$
r=1+3\frac{\dot H}{H^2}+\frac{\ddot H}{H^3},\,\,\,s= - \Big(\frac{2}{3 H}\Big)\,\frac{3 H \dot H+\ddot H}{3 H^2+2 \dot H}\,.
$$
With the help of (\ref{8}) and (\ref{16}), one can find that
\begin{equation}\label{19}
r=1+\frac{mn(n+1)\,\tau^{\displaystyle n}}{\Big(mn+\tau^{\displaystyle n+1}\Big)^3}\Big[(n+2)\tau^{\displaystyle n}-3\Big(mn+\tau^{\displaystyle n+1}\Big)\Big],
\end{equation}
\begin{eqnarray}\label{20}
s=\frac{2mn(n+1)\,\tau^{\displaystyle n}}{3\left[2mn(n+1)\,\tau^{\displaystyle n}-3\Big(m n+\tau^{\displaystyle n+1}\Big)^2\right]}\nonumber\\
\times \frac{(n+2)\tau^{\displaystyle n}-3\Big(m n+\tau^{\displaystyle n+1}\Big)}{m n+\tau^{\displaystyle n+1}}.
\end{eqnarray}

As can be easily seen from equations (\ref{19}) and (\ref{20}), our model is able to appear in the $\Lambda$CDM point $\{r, s\} = \{1, 0\}$ at the beginning moment $\tau=0$, at a final time $\tau \to \infty$, and also many times at the certain intermediate instants, determined by the roots of the following equation
$$
3\Big(m n+\tau^{\displaystyle n+1}\Big)^2=2mn(n+1)\,\tau^{\displaystyle n}.
$$
The final stage of cosmic evolution on $\{r,s\}$ plane is shown in Fig. 4, for $n=1/2$ and $1$.
\begin{figure}[thbp]
\centering
\includegraphics[width=0.4\textwidth,height=0.35\textwidth]{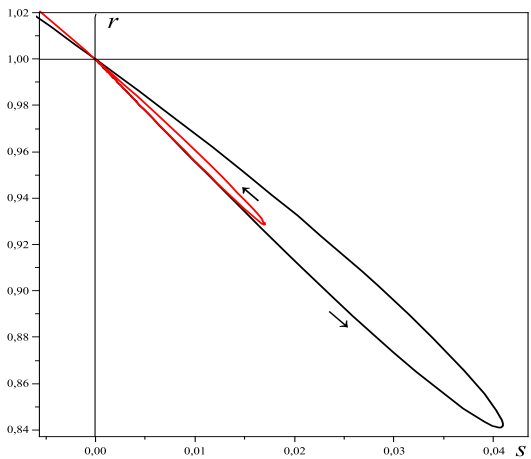}
\caption{The lines on $\{r, s\}$ plan for different values of parameter $n=1/2,\,1$ corresponding to the cases represented in Fig. 1. The $\{1, 0\}$ point is related to the location of standard $\Lambda\mbox{CDM}$ model.}
\label{Figure_4}
\end{figure}

\subsection{\normalsize{ Om diagnostic}}

Recently, a new cosmological parameter named $Om$  was proposed \cite{Sahni,Tong}.  It is a combination of the Hubble parameter and the cosmological
redshift   and provides a null test of dark energy. As a complementary to $\{r, s\}$, this diagnostic helps to distinguish the present matter density in different models more effectively.  $Om$ diagnostic has been discussed together with statefinder for many cosmological models of dark energy. It was
introduced to differentiate $\Lambda$CDM from other dark energy
models. For $\Lambda$CDM model, $Om = \Omega_{m0}$ is a constant, independent
of redshift $z$. The main utility for $Om$ diagnostic is that the quantity
of $Om$ can distinguish dark energy models with less dependence
on matter density $\Omega_{m0}$ relative to the equation of state of dark energy. For this end, we express the scale factor $a$  in terms of redshift $z$ by the relation
\begin{equation}\label{21}
a =\frac{a_0}{1+z},
\end{equation}
where  $a_0$ is the present value of scale factor. The starting point is the Hubble parameter which is used to determine the $Om$ diagnostic as follows:
\begin{equation}\label{22}
Om(x) = \frac{h^2(x)-1}{x^3-1},
\end{equation}
where $x=1+z, \,h(x)=H(x)/H_0$. \begin{figure}[thbp]
\centering
\includegraphics[width=0.4\textwidth,height=0.35\textwidth]{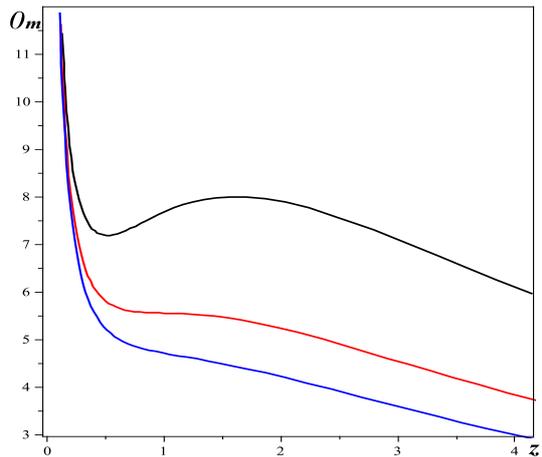}
\caption{The $Om$ parameter is plotted against the redshift $z$. Here, we use $n=1$ and $m=m_1=27/256$ (black line), $m=1.3 m_1$ (red line), and $m=1.5 m_1$ (blue line).}
\label{Figure_5}
\end{figure}
From the equations (\ref{8}), (\ref{9}) and the definition (\ref{22}), we can conclude that the explicit dependence $h(z)$ can be established for $n=1$ only. In this case, we have
$$
\tau = \frac{1}{2}\Big[\sqrt{\ln^2(x)+4m}-\ln(x)\Big],
$$
and
\begin{equation}\label{23}
h(x)=1+\frac{4m}{\Big[\sqrt{\ln^2(x)+4m}-\ln(x)\Big]^2}.
\end{equation}

$$
x(\tau)=\exp\Big(\displaystyle \frac{m}{\tau^{\displaystyle n}}-\tau\Big).
$$

Nevertheless, one can plot the $Om$ diagnostic parameter on $\{Om,x\}$ plane using the parametric representation as follows
$$
Om(\tau)=\frac{\Big(\displaystyle 1+\frac{m n}{\tau^{\displaystyle n+1}}\Big)^2-1}{\exp\Big(\displaystyle \frac{3m}{\tau^{\displaystyle n}}-3\tau\Big)-1},
$$
We find from Fig. 5 that the $Om(z)$ with $m=27/256$ is of negative slope almost everywhere, except for a certain interval of $z$. With $m$ increasing, the $Om(z)$ becomes of negative slope for all $z$.  According to \cite{Sahni}, it should suggest quintessence like behavior of our model ($w > -1$). This allows us to reconstruct this model  by means of a standard scalar field in Section 4.

\subsection{\normalsize{Supernovae type Ia constrains}}

As known, the SN Ia Union2 database includes 557 SNIa \cite{Liao} and provides one of the possible observational restrictions on the cosmological models [28--30].
One should follow the maximum-likelihood approach under which one minimizes $\chi^2$ and hence measures the deviations of the theoretical predictions from the observations.
Since SN Ia behave as excellent standard candles, they can be used to directly measure the expansion rate of the Universe upto high redshift, comparing with the present rate. The SN Ia data gives us the distance modulus $\mu$ to each supernova.
In the flat universe,the theoretical distance modulus is given by
\begin{equation}\label{24}
\mu_{th}(p_s;z) = 5 \log_{\,10} (d_L/Mpc) +25
\end{equation}
where $d_L$ is the luminosity distance, and $p_s$ denotes model parameters.  For theoretical calculations,
the luminosity distance $d_L$ of SNe Ia is defined as follows
\begin{equation}\label{25}
d_L = (1+z)\int\limits_o^z \frac{d z'}{H_0 E(z')},
\end{equation}
where $E(z)=H(z)/H_0$ or $E(z)\equiv h(z)$.

We choose the marginalized nuisance parameter \cite{Nesseris} for $\chi^2$
$$
\chi^2_{SNe} = A -\frac{B^2}{C}
$$
where
$$
A =\sum_{i}^{557}\Big[ \frac{\mu_{obs}(z_i)-\mu_{th}(p_s;z_i)}{\sigma_i}\Big]^2,
$$
$$
B =\sum_{i}^{557}\frac{[\mu_{obs}(z_i)-\mu_{th}(p_s;z_i)]}{\sigma_i^2},
$$
$$
C=\sum_{i}^{557}\frac{1}{\sigma_i^2},
$$
and $\sigma_i$ is the 1$\sigma$ uncertainty of the observed data $\mu_{obs}(z_i)$, and $p_s = \{H_0,\,m,\,n\}$ is the set of free parameters of the model to be defined through the best fit analysis.

Keeping in mind that we are dealing with the toy model, let us restrict ourself with the case $n=1$, when we are able to follow an analytic calculation up to a certain point. In this case,  there are only two  parameters to be assessed: $m$ and $H_0$.

By substituting (\ref{22}), we can represent the  luminosity distance (\ref{25}) as
\begin{eqnarray}
d_L = \frac{(1+z)}{H_0}\left(z-4m\times \phantom{.\frac{\sqrt{\frac{A}{B}}{\frac{a}{b} B}}.} \right.\nonumber\\
\left.\times \int\limits_o^z \frac{d z'}{4m + \Big[\sqrt{\ln^2(1+z')+4m}-\ln(1+z')\Big]^2}\right).\nonumber
\end{eqnarray}
Let us rewrite this expression approximately for a small value $z$ compared with unity, that is for the epoch close to the present time. Retaining in the series expansion of the power and logarithmic functions only linear terms, one can obtain
\begin{equation}\label{26}
d_L \approx \frac{z(1+z)}{2 H_0}\left(1+\frac{1-2\sqrt{m}}{12 m}\, z^2\right).
\end{equation}
Substituting this expression into (\ref{24}), we obtain  the theoretical distance modulus  in terms of the redshift parameter $z$.

Following the usual procedure (see, e.g. \cite{Liao} and bibliography therein), we can find the best-fit values of $m$ and $H_0$. As we use the approximation of $z\ll 1$, it is reasonable to take into account the restricted data on SNe Ia, presented in \cite{Riess2} and shown in Fig. 6. There, we ignore the confidence intervals of the data for the sake of simplicity.

With the aim of a preliminary study of our model for only the case $n=1$ , we have to roughly estimate its parameters. In the best-fit procedure mentioned above, we use the observational errors  at the 1$\sigma$ level and the data for SNeIa with $z<0.1$. One could find that the best fit parameters are $m=0.2025$ and $H_0=0.00011$. As can be seen, the curve $\mu_{th}(z)$ essentially deviates from the observational data at $z_i$ close to unity. Obviously, this is the consequence of approximation which we made.
\begin{figure}[thbp]
\centering
\includegraphics[width=0.4\textwidth]{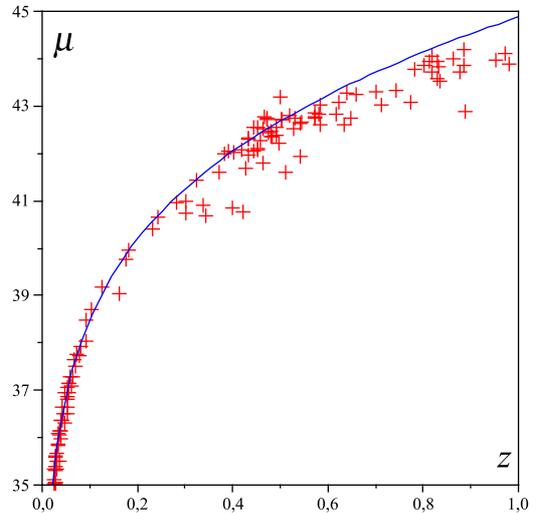}
\caption{The best-fitted distance modulus $\mu(z)$ based on approximation (\ref{26}) is plotted as function of redshift.}
\label{Figure_6}
\end{figure}

\section{\Large{Scalar-Field Reconstruction}}

The question, that naturally arises,  is as follows. How could this model be realized with some well-known forms of matter or field? The matter content of our Universe is complex and quite diverse.  Nevertheless, let us keep in mind that we are dealing here with the toy model. It allows us to attempt to build this model with a single source of gravity.

We assume now that the origin of the gravity in the universe is a single  self-interacting scalar field $\phi$ which couples minimally to gravity. The energy density $\rho$ and pressure $p$ for a scalar field with potential $V(\phi)$ are given by
\begin{equation}\label{27}
\rho=\frac{1}{2} \dot \phi^2 + V(\phi),\,\,\,\,p= \frac{1}{2} \dot \phi^2 - V(\phi),
\end{equation}
respectively. Scalar field evolution is governed by the equation of motion
$$
\ddot \phi + 3 H \dot \phi + \frac{d\, V}{d\,\phi} = 0,
$$
which can be easily obtained from (\ref{5}) by means of substitution (\ref{21}).

Combining two equations in (\ref{27}) and taking into account (\ref{10}) and (\ref{11}), we can obtain the following equations
\begin{equation}\label{28}
\phi\, '\, ^2= \frac{2 m n (n+1)}{\tau^{\displaystyle n+2}},
\end{equation}
\begin{equation}\label{29}
V(\tau)= 6 H_0^2\left(1+\frac{mn}{\tau^{\displaystyle n+1}}\right)^2-\frac{2 m n (n+1)H_0^2}{\tau^{\displaystyle n+2}},
\end{equation}
where a prime stands for  differentiation with respect to the dimensionless  cosmic time $\tau$.
\begin{figure}[thbp]
\centering
\includegraphics[width=0.45\textwidth]{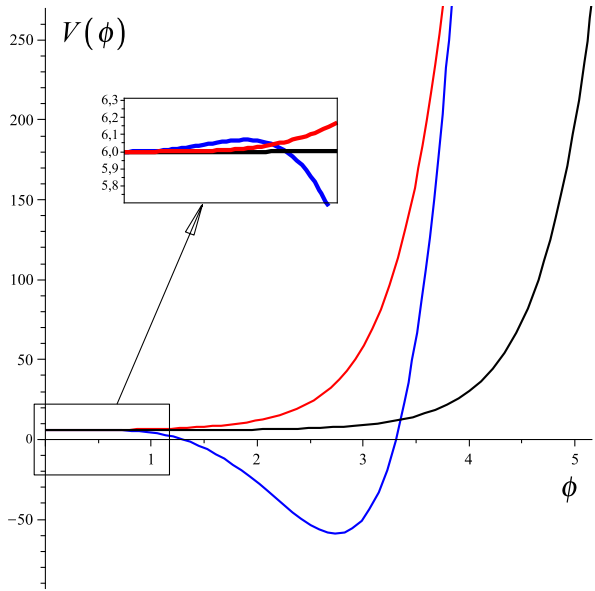}
\caption{Shows the potential versus field $\phi$. We use the same convention on values of $n$ and $m$ as in Fig. 1, and $H_0=1$. The enlarged piece of the main panel is shown in the embedded panel.}
\label{Figure_7}
\end{figure}

Integrating equation (\ref{28}), we can readily get
\begin{equation}\label{30}
\phi(\tau)=\pm \sqrt{\frac{8m(n+1)}{n}}\,\tau^{\displaystyle -n/2} + \phi_0,
\end{equation}
where $\phi_0$ is a constant of integration. Substituting $\tau$ from (\ref{30}) into (\ref{29}), one can reconstruct potential in the following form
\begin{eqnarray}
V(\phi)= 6 H_0^2\left[1+m n A_{mn}{(\phi-\phi_0)^{\displaystyle \frac{2(n+1)}{n}}}\right]^2 \nonumber\\
-\frac{n^2 H_0^2}{4} A_{mn}^{ \displaystyle (n+2)/(n+1)}{(\phi-\phi_0)^{\displaystyle \frac{2(n+2)}{n}}},\label{31}
\end{eqnarray}
where $A_{mn}=[n/8m(n+1)]^{ \displaystyle (n+1)/n}$. As an example, we plot the graphs of potential  (\ref{31}) in Fig. 7 for the cases: a) $n=1/2,\,m=(25/96)\sqrt{5/3}$, b) $n=1,\,m=27/256$, and c)  $n=2,\,m=1/27$.

\section{\Large{Conclusions}}

Thus, we have proposed and briefly investigated the toy model of complete cosmic history. It is hard to expect a full description of all  complex processes occurring in the universe in the framework of such a simple model. Our intention is to demonstrate that there are models that could cover, at some approximation, a wide period of cosmic evolution (cf., e.g. \cite{Perico}). At the same time, this model itself is the analytically accurate one  and therefore very suitable for the deeper studying of the main properties of cosmic expansion. One could express the hope for further development of this model in terms of  some modification of the sources and the evolutionary law of the Hubble parameter. At least, we believe this model merits a deeper study.

\noindent\hrulefill

\end{document}